\documentstyle[epsfig,longtable]{aipproc}

\begin{document}
\title{The TRIUMF \\ Parity Violation Experiment}

\author{
A.R.~Berdoz$^{b}$, J.~Birchall$^{e}$, J.D.~Bowman$^{d}$, 
J.R.~Campbell$^{e}$, C.A.~Davis$^{e,f}$, A.A.~Green$^{g}$,
P.W.~Green$^{a}$, A.A.~Hamian$^{e}$, D.C.~Healey$^{f}$, R.~Helmer$^{f}$,
S.~Kadantsev$^{c}$, Y.~Kuznetsov$^{c}$, R.~Laxdal$^{f}$, L.~Lee$^{e}$,
C.D.P.~Levy$^{f}$,  R.E.~Mischke$^{d}$, S.A.~Page$^{e}$,
W.D.~Ramsay$^{e}$, S.D.~Reitzner$^{e}$,  G.~Roy$^{a}$,
P.~Schmor$^{f}$, A.M.~Sekulovich$^{e}$, J.~Soukup$^{a}$, 
G.M.~Stinson$^{a}$, T.J.~Stocki$^{a}$, V.~Sum$^{e}$, N.~Titov$^{c}$, 
W.T.H~van~Oers$^{e}$, R.J.~Woo$^{e}$, A.~Zelenski$^{c,f}$ \\
~~\\ (talk by W.D. Ramsay at CIPANP97)\\}
\address{                
                       (E497 Collaboration)\\
$^{a}$University of Alberta, Edmonton, Alberta T6G 2N5, Canada\\
$^{b}$Carnegie Mellon University, Pittsburgh, PA 15213, USA\\
$^{c}$Institute for Nuclear Research, 117312 Moscow, Russia\\ 
$^{d}$Los Alamos National Laboratory, Los Alamos, NM 87545 USA\\
$^{e}$University of Manitoba, Winnipeg, Manitoba, R3T 2N2, Canada\\
$^{f}$TRIUMF, Vancouver, British Columbia, V6T 2A3, Canada\\
$^{g}$University of the Western Cape, Bellville 7535, South Africa\\ }

\maketitle

\begin{abstract}
 
An experiment (E497) is underway at TRIUMF to measure the 
angle-integrated, parity violating longitudinal analyzing power,
$A_z$, in proton-proton elastic scattering, to a precision of $\pm 0.2
\times 10^{-7}$. The experiment uses a 221 MeV longitudinally
polarized proton beam incident on a 40 cm liquid hydrogen target. The
beam energy is carefully chosen so that the contribution to $A_z$ from
the J=0 parity mixed partial wave ($^1\!S_0-^3\!P_0$) integrates to
zero over the acceptance of the apparatus, leaving the experiment
sensitive mainly ($>95\%$) to $A_z$ arising from the
$^3\!P_2-^1\!D_2$, J=2 wave. To  minimize sources of systematic error,
the \mbox{TRIUMF} ion source and cyclotron parameters have been
refined to the extent that helicity correlated beam changes are at an
extremely low level, and specialized instrumentation on the E497
beamline is able to measure residual helicity correlated modulations
to a precision consistent with the goals of the experiment.  A data
taking run in February-March, 1997 logged approximately 12\% of the
desired data and produced a preliminary result,  $A_z = (1.1 \pm 0.4
\pm 0.4) \times 10^{-7}$, where the error is statistical only. 

\end{abstract}

\section*{Introduction}

\begin{figure}[b!]
\centerline{\epsfig{file=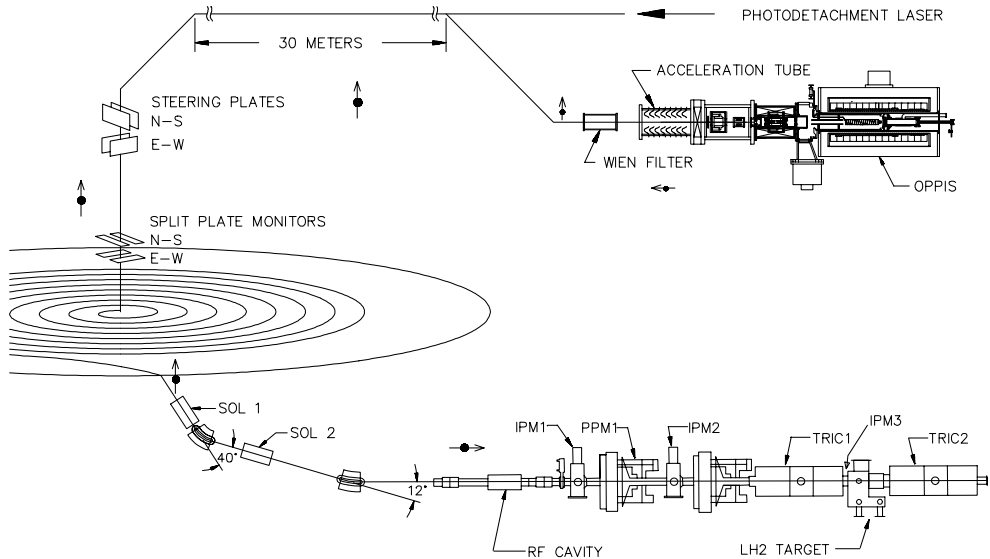, height=13.0cm, angle=90}}
\vspace{10pt}
\caption{General layout of the TRIUMF parity experiment.  (OPPIS:
Optically Pumped Polarized Ion Source;  SOL: Spin Precession Solenoid;
IPM: Intensity Profile Monitor; PPM: Polarization Profile Monitor;
TRIC: Transverse Field Ionization Chamber)}
\label{parityfg_wdr}
\end{figure}

Figure \ref{parityfg_wdr} shows the overall layout of TRIUMF
experiment E497. The polarized beam is prepared in an optically pumped
polarized ion source, the spin is precessed into the vertical
direction by a Wien filter, and the beam is accelerated to 221 MeV in
the TRIUMF cyclotron.  The extracted beam current is 200 nA and the
polarization is typically 80\%. The beamline precesses the spin into
the longitudinal direction and transports the beam to the parity
experimental area, where it passes through a series of beam diagnostic
and control devices before it is scattered from a 40 cm thick liquid
hydrogen target. Hydrogen filled transverse electric field ionization
chambers measure the current before and after the target.
Approximately 4\% of the incident protons scatter due to the strong
nuclear force between the incident and target protons.  However,
because of the simultaneous presence of the weak nuclear force, the
scattering fraction is expected to be enhanced very slightly, by about
one part in $10^7$, if the incident proton spin is aligned with the
beam direction, and reduced by the same fraction if the proton spin is
opposite to the beam direction. This difference is expressed as the
parity violating longitudinal analyzing power,  $A_z = (\sigma^+ -
\sigma^-)/(\sigma^+ + \sigma^-)$, where $\sigma^+$ and $\sigma^-$ are
the scattering cross sections for positive and negative helicity.  The
goal of E497 is to measure $A_z$ with a precision of $\pm 0.2 \times
10^{-7}$.

\subsection*{Choice of Energy}

\begin{figure}[h!] 
\centerline{\epsfig{file=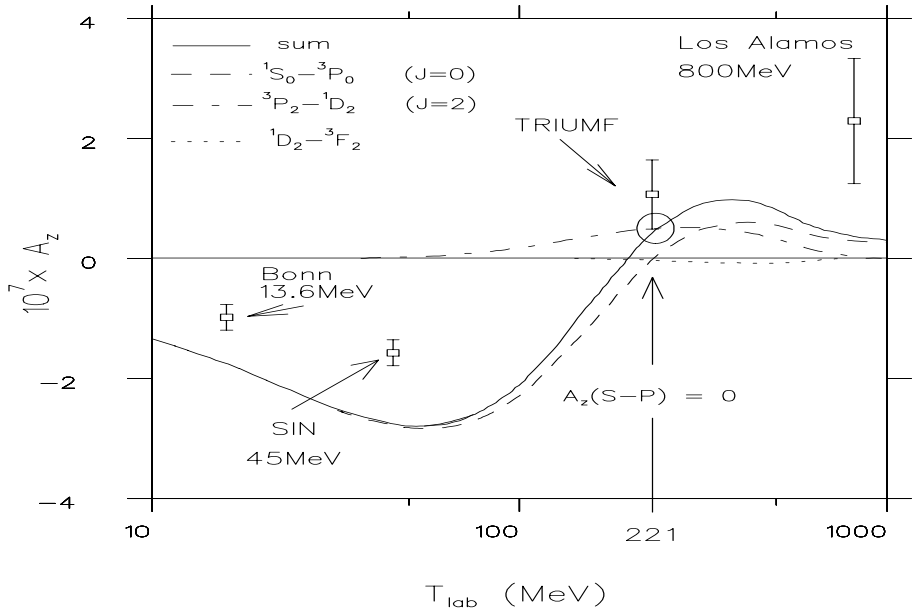, height=8.0cm, width=12.0cm}} 
\vspace{10pt} 
\caption{Partial wave contributions to $A_z$. The curves are from 
Driscoll and Miller[1]. Also shown are data from Bonn[5], SIN[6],
Los Alamos[7], and the TRIUMF preliminary result.}
\label{drismill} 
\end{figure}

Figure \ref{drismill} shows the results of meson exchange 
calculations by Driscoll and Miller\cite{drismill89} using the Bonn
meson exchange potential\cite{machleidt87,machleidt89} for the strong
interaction and the DDH\cite{ddh80} predictions for the weak coupling
parameters. The calculated $A_z$ is shown broken down into
contributions from the various parity mixed partial waves.  

Because the variation of $A_z$ with angle is determined only by the
well-known strong interaction, one can calculate the energy at which
the lowest order, J=0, parity mixed partial wave ($^1\!S_0-^3\!P_0$)
will integrate to zero over the acceptance of the detectors.   The 221
MeV energy of the TRIUMF experiment is chosen so that the measured
$A_z$ comes  exclusively\footnote{At this energy the contribution to
$A_z$ from the $^1\!D_2-^3\!F_2$ partial wave is only 5\% of that from
$^3\!P_2-^1\!D_2$.} from the J=2 parity mixed partial wave
($^3\!P_2-^1\!D_2$) which, in the meson exchange model, comes
from $\rho$-meson exchange.  Theoretical predictions 
\cite{drismill89,simonius88,drismill89-2,grachshmat93,iqbalnisk94}
for $A_z$ at this energy, span a substantial range.  It is hoped that
the TRIUMF measurement, in selecting the effects of only one partial
wave, will provide a definitive result.
\section*{Systematic Errors}

If beam properties other than the spin direction change when the spin
is flipped, it can affect the measured $A_z$.  {\em Random} changes
will appear as noise, and simply increase the time required to achieve
a given statistical precision; {\em coherent} changes, that is
changes which are synchronized with spin reversal, appear as a false
$A_z$. Table 1 summarizes the corrections to the February-March 1997
data for all sources of systematic error that are measurable with the
parity apparatus. Notice that the corrections are very small and, in
most cases are zero to within statistics. This is the result of an
exhaustive program of reducing all unwanted helicity correlated
modulations to a bare minimum and of minimizing the sensitivity to
these modulations.

\subsection*{Transverse Polarization Components}

\begin{table}[b!]
\caption{Corrections ($\Delta A_z$) to the February-March 1997 Data.}
\label{table1}
\begin{tabular}{lcc}
   Item         & $10^{7}(\Delta A_z$)&         Comment                \\
\tableline
         $P_x$  &   0.01 $\pm$ 0.09  & Correction very small          \\
         $P_y$  &   0.02 $\pm$ 0.20  & for all data sets              \\
\tableline
         $yP_x$ & -0.001 $\pm$ 0.002 & sensitivity extracted          \\
         $xP_y$ &   0.00 $\pm$ 0.02  & from real data correlations    \\
\tableline
 $\Delta\sigma$ &   -0.07 $\pm$ 0.20 & sensitivity from               \\
                &                    & separate measurement           \\
\tableline
 $\Delta x$     &   -0.03 $\pm$ 0.15 & correction $\sim0$ for all sets\\
\tableline
 $\Delta y$     &    0.23 $\pm$ 0.16 & some cancellation between sets \\
\tableline
 $\Delta I/I$   &    0.05 $\pm$ 0.05 & using interleaved CIM data      \\
\tableline
   Total        &     0.2 $\pm$ 0.4  &                                \\
\tableline
\end{tabular}
\end{table}

Ideally, the polarization is purely longitudinal.  Unwanted transverse
components, $P_t$ = $P_x$ or $P_y$, are kept small ($<0.001$) but
couple to the large parity allowed analyzing power and, when combined
with an off-center beam, generate a false $A_z$. The sensitivity to
transverse components is found by setting the spin precession magnets
for large $P_t$ and measuring the false $A_z$ as a function of beam
position.  This procedure also determines the ``polarization neutral
axis'' -- the beam position with minimum sensitivity to transverse
polarization components. A beam position servo system then holds the
average beam position on this axis to within about 50$\mu m$. During
data taking, the polarization profile monitors (PPM1 and PPM2 in
figure \ref{parityfg_wdr}) continuously measure the transverse
components and a correction is applied.

Even if the average transverse polarization is zero, the first moment
of transverse polarization need not be.  For example, the polarization
may be up on the left of the beam and down on the right.  Such
unwanted polarization profiles also cause false $A_z$.  For this
reason, the first moments of transverse polarization, $<\!xP_y\!>$ and
$<\!yP_x\!>$ are continuously monitored by the PPMs.  They show a
random variation from run to run, but typical values are a few $\mu m$ 
for a one hour run, averaging to near zero over a 20 to 30 hour data
set.  In a drift space, first moments vary linearly with position
along the beamline, making it possible to adjust the beam optics so
that the first moments pass through zero at a point which minimizes
their effect. The sensitivity to intrinsic first moments must be
determined by looking at correlations between apparent $A_z$ and the
$<\!xP_y\!>$ and $<\!yP_x\!>$ measured by the PPMs. For the February,
1997 data, this sensitivity was consistent with zero.

\subsection*{Beam Size Modulation}

Because the beam is different upstream and downstream of the liquid
hydrogen target, a beam size change affects the current differently in
the upstream and downstream detectors (TRIC1 and TRIC2 in figure
\ref{parityfg_wdr}).  Actual coherent size modulation is typically
only a few tenths of one $\mu m$ on a $\sigma = 5 mm$ beam, but it can
cause a detectable shift in the measured $A_z$. To measure the
sensitivity to beam size change, a relatively large coherent beam size
modulation is introduced using a pair of fast ferrite-cored quadrupole
magnets. The sensitivity measured in this way is then multiplied by
the actual coherent size change measured during data taking to obtain
the correction.

\subsection*{Beam Intensity Modulation}

Coherent intensity modulation from the optically pumped polarized ion
source is very small, usually only a few parts in $10^5$. The coherent
intensity modulation is measured constantly during data taking.  In
addition, the sensitivity is monitored by interleaving with the normal
data a 10\% subset with artificially enhanced ($\sim 0.1\%$) coherent
intensity modulation (CIM). 

\subsection*{Beam Position Modulation}

If the beam position changes with spin reversal then a false $A_z$
is introduced proportional to the magnitude of the coherent change and
the mean position of the beam.  The farther the beam is off-axis, the
greater the sensitivity to coherent position change.  Sensitivity to
position modulation is measured by introducing relatively large
position modulation using the same fast ferrite-cored steering magnets
which are used for the beam position servo. This information is then
combined with the actual beam position information recorded during the
run to obtain the correction.  

\subsection*{Overall Bias}

Although corrections for all systematic errors measurable with the
E497 apparatus appear to be well under control, the possibility
remains that some additional effect could be present which  cannot be
measured, but which will cause a false $A_z$.  One example is energy
modulation. Such effects must be handled by reversing the sign of the
spin in the beamline relative to that at the ion source or cyclotron. 
The real asymmetry from $A_z$ will then reverse sign, but the false
effect will not, but will simply appear as an overall bias on the
measured $A_z$. For this reason, runs were made under four
configurations representing all possible combinations of spin
direction at the ion source, cyclotron, and  parity apparatus (see
figure \ref{parityfg_wdr}). Only four different conditions are needed
because changing the spin direction in the ion source is the same as
reversing the definition of spin ``up''. 

\begin{figure}[htb]
\centerline{\epsfig{file=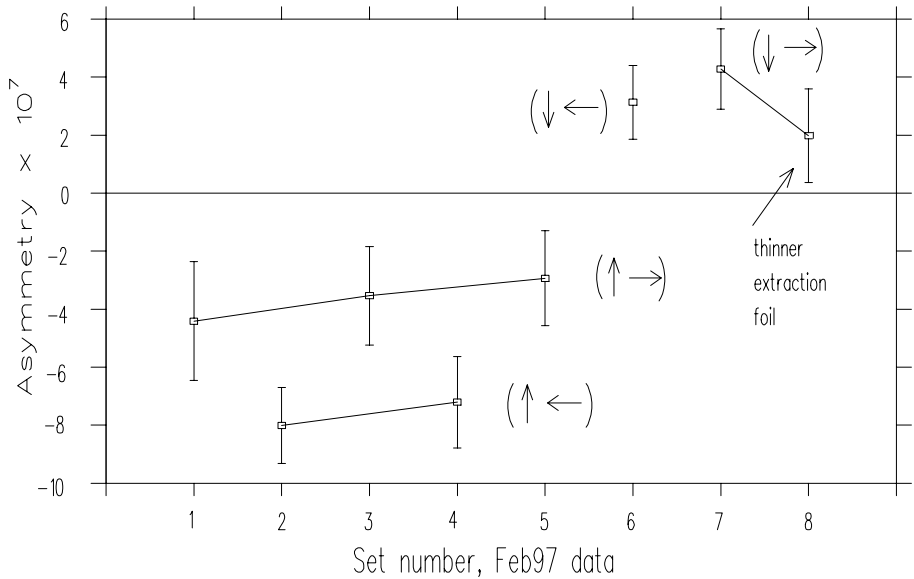, width=12.0cm}}
\vspace{10pt}
\caption{Data from the February-March, 1997 run after correction for
all measurable sources of systematic error.  In brackets, the
left-hand arrow shows the orientation of the spin in the cyclotron and
the right-hand arrow shows the direction at the parity apparatus.  The
sign of the plotted asymmetry is such that it is equal to $+A_z$ for
positive helicity at the apparatus and equal to $-A_z$ for negative
helicity at the apparatus. (A right arrow indicates positive
helicity.)}
\label{eps_feb97}
\end{figure}

\section*{Results of February-March, 1997 Run}

A total of 231 runs of ``real data'' were taken during February and
March.  80\% was polarized, 10\% was unpolarized and 10\% was
unpolarized with intentional coherent intensity modulation.  105 hours
of polarized data passed all the cuts. Figure \ref{eps_feb97} shows
the results from these data. Each point has already been corrected for
the effects of coherent change in beam intensity, position, size,
transverse polarization, and first moments of transverse polarization.
The asymmetry apparently contains a part which is related to spin in
the cyclotron and a part which is related to spin in the parity
apparatus.  To extract $A_z$, it is assumed that only that part of the
asymmetry which reverses with spin at the parity apparatus is true
$A_z$. In principle, it is enough to reverse the beamline helicity.  If
this is done fairly frequently, it will establish the constancy of the
offset and the true $A_z$ can be extracted.  The reversal of the spin
in the cyclotron was done in an attempt to locate the source of the
offset.

The last point, set 8,  on figure \ref{eps_feb97} requires some
explanation.  It was taken with the normal $6\,mg/cm^2$ stripping foil
replaced with a $2.5\,mg/cm^2$ foil. This was done to test the theory
that the bias might be arising from some interaction with the
stripping foil.  It appears the bias was reduced by the thin foil, but
unfortunately the variation could be statistical. For the Parity 
data-taking run starting in August, 1997, The experiment will use a
very thin, $200 \mu g/cm^2$ stripping foil.  This should make clear
whether or not the large $A_z$ offset arises from some interaction
with the stripping foil.

The raw $A_z$ from February-March, 1997 is $(1.3\pm0.4)\times 10^{-7}$
and, after all corrections are applied, the final $A_z$ is
$(1.1\pm0.4\pm0.4)\times 10^{-7}$, where the first error comes from
statistical uncertainty in the raw $A_z$ and the second error is
dominated by statistical uncertainty in measurement of the helicity
correlated quantities. Since the uncertainty in the correction is also
statistical, the two errors can be combined, giving  $A_z = (1.1 \pm
0.6) \times 10^{-7}$.  We consider this number very preliminary
because of the relatively large systematic shift which had to be
removed by averaging over different beamline helicities.

\end{document}